Jean Bricmont
FYMA, 2, ch. du Cyclotron
B-1348 Louvain-la-neuve, Belgium
Tel:32-10-473277
Fax: 32-10-472414

\catcode`\@=12 {\count255=\time\divide\count255 by 60
\xdef\hourmin{\number\count255} \multiply\count255 by-60\advance\count255
by\time \xdef\hourmin{\hourmin:\ifnum\count255<10 0\fi\the\count255}}
\def\ps@draft{\let\@mkboth\@gobbletwo \def\@oddhead{} \def\@oddfoot {\hbox
to
7 cm{$\caliptstyle Draft\ version:\ \draftdate$ \hfil} \hskip
-7cm\hfil\rm\thepage \hfil} \def\@evenhead{}\let\@evenfoot\@oddfoot}
\catcode`\@=12 \def\draftdate{\number\month/\number\day/\number\year\ \ \
\hourmin } \def\draft{\pagestyle{draft}\thispagestyle{draft}}
\documentstyle[12pt]{article} 
 \newcommand{\be}{\begin{eqnarray}}
\newcommand{\en}{\end{eqnarray}} 
\newcommand{\no}{\noindent} \newcommand{\vs}{\vspace}
  
  \newcommand{\Bbb}{\bf}
\newcommand{\p}{\partial} 
 \newcommand{\var}{\varphi} 

\title{{\bf Coupled Analytic Maps}}

\author{J.Bricmont\thanks{Supported by EC grants SC1-CT91-0695 and
CHRX-CT93-0411}\\ UCL, Physique Th\'eorique, B-1348, Louvain-la-Neuve,
Belgium\and A.Kupiainen\thanks{Supported by NSF grant DMS-9205296} \\
Helsinki University, Department of Mathematics,\\ Helsinki 00014, Finland}

\date{}

\begin{document}

\maketitle \begin{abstract}

We consider a lattice of weakly coupled expanding circle maps. We
construct,
via a cluster expansion of the Perron-Frobenius operator, an invariant
measure
for these infinite dimensional dynamical systems which exhibits
space-time-chaos. \end{abstract}

\section{Introduction}

\draft

It is an important problem to determine which parts of the rich theory of
finite dimensional dynamical systems ( e.g. hyperbolic attractors, SRB
measures \cite{ER}) can be extended to infinite dimensional ones. The
latter
are usually given by non-linear partial differential equations of the form
$\p_t u = F(u, \p u, \p^2 u,\dots)$, i.e. the time derivative of $u(x,t)$
is
given in terms of $u(x,t)$ and its partial space derivatives. One would
like
to find natural invariant measures for the flow.

In a bounded spatial domain and $F$ suitably dissipative, such equations
tend
to  have finite dimensional attracting sets \cite{Te} and thus fall in into
the class  of finite dimensional systems. Genuinely infinite dimensional
phenomena are  expected to occur for dissipative PDE's on unbounded domains
\cite{ch}.  In particular, invariant measures for the flow might have
infinite dimensional  supports and there might be several of them
(corresponding to a ``phase transition").

A  class of dynamical systems, possibly modelling such PDE's, are obtained
by
discretizing space and time and considering a recursion \be u(x,t+1) =
F(x,u(\cdot,t)) \en i.e. $u(x,t+1)$, with $x$ being a site of a lattice, is
determined by the values taken by $u$ at time $t$ (usually on the sites in
a
neighbourhood of $x$). For a suitable class of $F$'s such dynamical systems
are called Coupled Map Lattices \cite{Ka,Ka2}.

The first rigorous results on such systems are due to Bunimovich and Sinai
$\;$ who studied a one dimensional lattice of weakly coupled maps
\cite{BS}.
They  established the existence of an invariant measure with exponential
decay of  correlations in space-time. Their method was to construct a
Markov
partition and to show uniqueness of the Gibbs state for the corresponding
two-dimensional spin system. This is a natural extension of the method used
for a single map or for hyperbolic systems \cite{Si,R}. These results were
strengthened by Volevich \cite{V} and extended by Pesin and Sinai $\;$ to
coupled hyperbolic attractors \cite{PS} (for a review, see \cite{BS2}). An
extension to lattices of any dimension is announced in \cite{V2} (for
coupled
hyperbolic attractors).

Since the Gibbs measure constructed by Bunimovich and Sinai describes
statistical  mechanics in two dimensions, the possibility of phase
transitions i.e.  non-uniqueness of invariant measure is open (for recent
results on this, see  \cite{Bl,Bl2,BC,Bu,MH,Po} and references therein). In
statistical mechanics, Gibbs measures are often easy to construct in weak
coupling (which corresponds to high temperature) and in strong coupling
(low
temperature)  using convergent expansions. The purpose of the present paper
is to develop these expansion methods for the dynamical system problems in
infinite dimensions.

We consider weakly coupled circle maps and derive a convergent cluster
expansion for the Perron-Frobenius operator (transfer matrix in the
statistical mechanics terminology). This allows us to prove exponential
mixing in space and time for an invariant SRB measure. These results are
similar to those of Bunimovich and Sinai, but our method works immediately
in
any dimension and is simpler. However, for technical reasons we need to
restrict ourselves to real analytic maps.

The Perron-Frobenius operator has been a powerful tool to analyze quite
general maps, of bounded variation \cite{LY} (for reviews see
\cite{Co,LM}).
This approach was used also for coupled maps in \cite{K}, but weaker
results
were obtained there in the infinite volume limit. An  open problem still
remains to develop expansion methods for coupled maps that are
 of bounded variation. These are the most natural candidates that might
exhibit interesting phase transitions as the coupling is increased.

We have tried to make the paper self-contained for readers having no
background in the expansion methods of statistical mechanics. Appendix 2
contains some of the standard combinatorical estimates needed. A reader who
is familiar with these methods will  find a slightly novel application of
them because our expansion is applied directly to the Perron-Frobenius
operator.

\section{Results}

\draft

We consider the following infinite dimensional dynamical system. The state
space of the system is $$ M = (S^1)^{{\bf Z}^d}, $$ the direct product of
circles over ${\Bbb Z}^d$, i.e., $m\in M$ is given as $m =
\{m_i\}_{i\in{\Bbb
Z}^d},\;m_i\in S^1$. $M$ carries the product topology and the Borel
$\sigma$-algebra inherited from $S^1$. To describe the dynamics, we
consider
a map $f: S^1\to S^1$ and let ${\cal F}: M\to M$ be $ {\cal F} =
\times_{i\in{\bf Z}^d}\;f $ i.e., ${\cal F}(m)_i = f(m_i)$. $\cal F$ is the
uncoupled map. The coupling map $\Phi: M\to M$ is given by \be \Phi(m)_i =
m_ie^{2\pi i\varepsilon \sum_j g_{|i - j|} (m_i, m_j)} \en where $g_n: S^1
\times S^1\to{\Bbb R}\;\;$ ($g$ will be chosen such that the sum converges)
and $\varepsilon>0$ is a parameter. We define now the coupled map $T: M\to
M$
by \be T = \Phi \circ \cal F. \en We assume the following:

\vs{2mm}

\noindent {\bf A}. $f$ is expanding and real analytic.

\vs{2mm}

\noindent {\bf B}. $g_n$ are exponentially decreasing and real analytic.

\vs{2mm}

More precisely, for B we assume that there exists a neighbourhood $V$ of
$S^1$ in ${\Bbb C}$ such that $g_n$ are analytic in $V\times V$ and \be
|g_n(u,v)|\leq C e^{-{\lambda}n} \en for some $C<\infty, {\lambda} > 0$ and
all $u,v\in V$.

\vs{2mm}

\noindent Let us denote by $\sigma_i$ the shifts in ${\Bbb Z}^d,\; i =
1,\dots,d$. Then our main results are

\vs{3mm}

\no {\bf Theorem}.{\it $\;$ Suppose $f$ and $g_n$ satisfy $A, B$. There is
an
$\varepsilon_0 > 0$ such that for $\varepsilon < \varepsilon_0$ there
exists
a Borel measure $\mu$ on $M$ such that

\vs{2mm}

\no 1. $\mu$ is invariant under $T$ and the shifts $\sigma_i$. For any
finite
$\Lambda$ $\subset \Bbb Z^d$, the marginal distribution of $\mu$ on
$(S^1)^\Lambda$ is absolutely continuous with respect to the Lebesgue
measure.

\vs{2mm}

\no 2. The ${\bf Z}^{d+1}$ action generated by $T$ and
$\{\sigma_i\}_{i=1}^{d}$ is exponentially mixing.

\vs{2mm}

\no 3. $T^n m\to\mu$ weakly as $n\to\infty$, where $m$ is the product of
Lebesgue measures on $S^1$.}

\vs{3mm}

\noindent {\bf Remark 1.} For a precise statement of 2, see Proposition 7.
Properties 1 and 2 are usually called ``space-time chaos".

\vs{2mm}

\noindent {\bf Remark 2.} The result holds for a much more general class of
interactions $\Phi :$ we could take $$ \Phi(m)_i = m_ie^{2\pi i\sum_{X:
i\in
X} g_X(m_X)} $$ where
 $X\subset {\Bbb Z}^d$, $m_X = m|_X, |X|$ is finite and $g_X:
(S^1)^{|X|}\to{\Bbb R}$ is analytic in $V^{|X|}$ with the bound $$
|g_X(z_X)|\leq \varepsilon^{|X|} e^{-{\lambda}\tau(X)} $$ where $\tau(X)$
is
the length of the shortest tree graph on the set $X$. Then the theorem
holds
again for $\varepsilon < \varepsilon_0$ small enough provided $g_{X+j} =
g_X,
j\in {\Bbb Z}^d$. Without this translation invariance $\mu$ is not
$\sigma_i$-invariant, but the other claims still hold. Similarily, $\cal F$
may be replaced by $\times_{i\in{\Bbb Z}^d}f_i$ where $f_i$ satisfy $A$,
uniformly in $i$. Our results also extend to
 coupled maps of the interval $[0,1]$, of the type considered by
Bunimovich
and Sinai \cite{BS}, provided we take their $f$ and $\alpha$ analytic.

\vs{2mm}

\noindent {\bf Remark 3.} We do not prove, but conjecture, that there is
only
one invariant measure whose local marginal distributions on the sets
$(S^1)^\Lambda, \Lambda \subset \Bbb Z^d, |\Lambda |$ finite are absolutely
continuous with respect to Lebesgue measure. This is true in a class of
measures with analytic marginals: the proof of 3 extends to the case where
m
is replaced with a measure satisfying some clustering and the analyticity
of
the densities of the local marginal distributions. This is similar to the
results of Volevich \cite{V,V2}. Thus $\mu$ can be considered as a natural
extension to the infinite dimensional context of an SRB measure.

\section{ Decoupling of the Perron-Frobenius operator}

Let $\Lambda\subset{\Bbb Z}^d$ be a finite connected set (a set $\Lambda$
in
${\Bbb Z}^d$ is connected if every point of $\Lambda$ has a nearest
neighbour
in $\Lambda$, or if $\Lambda$ consists of a single point). We denote $
M_\Lambda = (S^1)^{\Lambda} $. Let now $f_\Lambda= \times_{i\in\Lambda} f$
and let $\Phi_\Lambda : M_\Lambda\to M_\Lambda$ be given by (2) where the
sum
is over $j$ in $\Lambda$. We first construct a $T_\Lambda =
\Phi_\Lambda\circ
f_\Lambda$ invariant measure $\mu_\Lambda$ on $M_\Lambda$ and later $\mu$
as
a suitable limit of $\mu_\Lambda$. The measures $\mu_\Lambda$ will be
constructed by means of the Perron-Frobenius operator of $T_\Lambda$. To
describe this, we first collect some straightforward facts about $f$ and
its
invariant measure.

\vs{2mm}

\noindent The Perron-Frobenius operator $P$ for $f$ on $L^1(S^1)$ is
defined
as usual by $$ \int (g\circ f) h\;dm = \int g P h\; dm $$ for $h\in
L^1(S^1),
g\in L^\infty(S^1)$ and $dm$ the Lebesgue measure (i.e., $dx$ in the
parametrization $m = e^{2\pi ix})$. We wish to consider $P$ on a smaller
space, namely $H_\rho$, the space of bounded holomorphic functions on the
annulus $A_\rho = \{1-\rho < |z| < 1+\rho\}$. $H_\rho$ is a Banach space in
the $\sup$ norm. The assumption $A$ for $f$ implies the following spectral
properties for $P$:

\vs{3mm}

\noindent {\bf Proposition 1.} {\it There are constants $\rho_0 > 0, \gamma
>
1$ such that

\vs{2mm}

\noindent a) $P: H_\rho\to H_{\gamma\rho}$ is continuous for all $\rho\leq
\rho_0$.

\vs{2mm}

\noindent b) $P: H_\rho \to H_\rho$ can be written as \be P = Q+R \en with
$Q$ a 1-dimensional projection operator: \be Qg = \left(\int_{S^1} g\,
dm\right) h\equiv \ell (g) h \en with $h\in H_{\rho_0}, h > 0$ on $S^1$,
$\int_{S^1} h\,dm = 1$ and \be \parallel \mid R^n \parallel \mid \leq C
\mu^n,\;\; QR = RQ = 0 \en for some $\mu < 1,\; C < \infty$ and all $n$; we
use $\parallel \mid \cdot \parallel \mid$ to denote operator norms.}

\vs{3mm}

\noindent {\bf Remark 1.} a) is a consequence of the expansiveness of $f$
and
shows that the operator $P$ improves the domain of analyticity, while b)
means that there is a unique absolutely continuous invariant measure for
$f$,
with density in $H_{\rho_0}$, and the rest of the spectrum of $P$ in
$H_{\rho_0}$ is strictly inside the unit disc. Since all this is rather
standard (see \cite{Co,LM,LY}), we defer the proof to Appendix 1.

\vs{2mm}

\no{\bf Remark 2.} Throughout the paper $C$ will denote a generic constant,
which may change from place to place, even in the same equation.

\vs{2mm}

\noindent To describe the Perron-Frobenius operator ${\cal P}_\Lambda$ for
the coupled map $T_\Lambda$, we introduce some notation. We denote ${\cal
H}^\Lambda_\rho = \otimes_{i\in\Lambda} H_\rho,\;P_\Lambda =
\otimes_{i\in\Lambda} P_i$ where $P_i$ acts on the $i$-th variable. Also,
we
denote by $d m_\Lambda$ the Lebesgue measure on $M_\Lambda$. Then \be {\cal
P}_\Lambda = \Phi^\ast_\Lambda P_\Lambda \en where \be (\Phi^\ast_\Lambda
G)
(m) = \det D\Phi^{-1}_\Lambda (m) G(\Phi^{-1}_\Lambda(m)) \en and thus \be
\int( G \circ T_\Lambda) H\;dm_\Lambda = \int G {\cal P}_\Lambda
H\;dm_\Lambda \en for $G\in L^\infty (M_\Lambda), H\in L^1 (M_\Lambda)$. We
have of course to show that (9) is well defined. Actually, the strategy of
our proof will be to first derive a ``cluster expansion'' \cite{GJS,GJ} for
$\Phi^\ast_\Lambda$ in terms of localized operators with good bounds on
norms. Then we shall construct the invariant measure in Section 4 by
studying
the limit $n \to \infty$ of ${\cal P}^n_\Lambda$; a cluster expansion for
${\cal P}^n_\Lambda$ will be obtained by combining (5) for $P_\Lambda$ and
(11) below for $\Phi^\ast_\Lambda$.

\vs{3mm}

\noindent {\bf Proposition 2.} {\it Let $\Phi_\Lambda$ given by} (2){\it
with
$i,j\in\Lambda$, and let $g_n$ satisfy} {\bf B}. {\it The following holds
uniformly in $\Lambda:$ There exists $\rho_1 > 0$ and $\varepsilon_0 > 0$
such that for $\varepsilon < \varepsilon_0$, $\Phi^\ast_\Lambda$ maps
${\cal
H}^\Lambda_{\rho_1}$ into ${\cal H}^\Lambda_{\rho_1-\delta}$ where we may
take $\delta\to 0$ as $\varepsilon\to 0$. Moreover \be \Phi^\ast_\Lambda =
\sum_{\cal Y} \otimes_{Y\in\cal Y}\delta_{Y} \otimes 1_{\Lambda\setminus
\cal
Y} \en where $\cal Y$ runs through sets of disjoint subsets of $\Lambda$,
$\Lambda\setminus{\cal Y}= \Lambda\setminus\cup Y$ and $1_Z$ denotes the
imbedding of ${\cal H}^Z_{\rho_1}$ into ${\cal H}^{Z}_{\rho_1-\delta}$. The
operators $\delta_{Y}: {\cal H}^{Y}_{\rho_1} \to{\cal
H}^{Y}_{\rho_1-\delta}$
are bounded, with \be \parallel \mid \delta_{Y} \parallel \mid \leq
\eta^{|Y|} \sum_\tau e^{-\frac{\lambda}{2} |\tau|} \equiv \eta^{|Y|} T (Y,
{\lambda}) \en where the sum runs over tree graphs on $Y$, $|\tau|$ is the
length of $\tau$ and $\eta\to 0$ as $\varepsilon\to 0$. }

\vs{3mm}

\noindent {\bf Proof.} Let us introduce decoupling parameters $s =
\{s_{ij}\}$, $i, j\in \Lambda$, $i < j$ in some linear order of $\Lambda$,
for our map $\Phi^\star_\Lambda$: \be (\Phi^\ast_\Lambda G)(m) = \det
D\Phi^{-1}_{\Lambda s} (m) G(\Phi^{-1}_{\Lambda s} (m))|_{s=1} \en where
\be
\Phi_{\Lambda s} (m)_i = m_ie^{2\pi i\varepsilon \sum_{j\in\Lambda} s_{ij}
g_{|i-j|} (m_i, m_j) } \en and we let $s_{ii}\equiv 1$, $s_{ij}=s_{ji}$.
Our
assumption {\bf B} for $g_n$ implies that there exists a $\rho_1 > 0$ such
that all $g_n$'s are holomorphic in $A_{\rho_1}\times A_{\rho_1}$ (
$A_\rho$
is the annulus) and (4) holds there, together with \be |\partial
g_n(u,v)|\leq C e^{-{\lambda}n}. \en where $\partial=\partial/\partial u$
or
$\partial/\partial v$. Consider now $s$ complex, in the polydisc $D_\Lambda
=
\{s_{ij}\vert i < j,\;|s_{ij}|< r e^{\frac{{\lambda}}{2} |i-j|} \} \subset
{\Bbb C}^{\frac12(|\Lambda|^2 - |\Lambda|)}$ (we shall take $r$ large for
$\varepsilon$ small). Then, we have the following

\vs{2mm}

\no {\bf Lemma}. {\it There exists $\varepsilon_0(\rho_1, r, {\lambda})$
such
that, for $\varepsilon < \varepsilon_0(\rho_1, r, {\lambda})
\;\;\Phi^{-1}_{\Lambda s}$ is a holomorphic family (for $s\in D_\Lambda$)
of
holomorphic diffeomorphisms from $A^\Lambda_{\rho_1-\delta}$ into
$A^\Lambda_{\rho_1}$ where $\delta\to 0$ as $\varepsilon\to 0$ (and
$A^\Lambda_\rho$ denotes the polyannulus $\{z_i\in A_\rho ,\;
i\in\Lambda\}$).

\no Moreover, the bound \be \|\det D\Phi^{-1}_{\Lambda s} \|\leq \exp
(C\varepsilon r (1+ {\lambda}^{-d}) |\Lambda|) \en holds uniformly in
$\Lambda$. $\| \cdot \|$ is the norm in} ${{\cal
H}^{\Lambda}_{\rho_1-\delta}}$.

\vs{2mm}

\no {\bf Proof}. (4) and (15) imply \be &&\big\vert D\Phi_{\Lambda,s}
(z)_{ij} - \delta_{ij}\big\vert\leq C\varepsilon r e^{-\frac \lambda2
|i-j|}
\\ &&\big\vert\Phi_{\Lambda,s} (z)_i - z_i \big\vert\leq C(1+
{\lambda}^{-d})\varepsilon r \en for $z\in A^\Lambda_{\rho_1}$ and $s \in
D_\Lambda$.  From these inequalities it follows easily that
$\Phi_{\Lambda,s}$ is a diffeomorphism from $A^\Lambda_{\rho_1}$ onto a set
containing $A^\Lambda_{\rho_1-\delta}$ provided $C(1+
{\lambda}^{-d})\varepsilon r < \delta$. The inverse of $\Phi_{\Lambda,s}$
then satisfies bounds (17), (18) too (with different constants) and then
(16)
follows from Hadamard's inequality \cite{DS}. \hfill$
\makebox[0mm]{\raisebox{0.5mm}[0mm][0mm]{\hspace*{5.6mm}$\sqcap$}}$ $
\sqcup$

\vs{2mm}

Thus, $\Phi^\ast_{\Lambda s}: {\cal H}^\Lambda_{\rho_1}\to {\cal
H}^\Lambda_{\rho_1-\delta}$ is a holomorphic family of operators, for $s\in
D_\Lambda$.

\vs{2mm}

The cluster expansion for $\Phi^\star_\Lambda$ is the following repeated
application of the fundamental theorem of calculus : \be \Phi^\ast_\Lambda
=
\Phi^\ast_{\Lambda s}|_{s=1} = \sum_\Gamma \int_{[0,1]^{|\Gamma|}}
\,ds_\Gamma\partial^\Gamma \Phi^\ast_{\Lambda s}|_{s_{\Gamma^c} = 0} \equiv
\sum_{\Gamma} \delta_{\Lambda\Gamma} \en where the notation is as follows.
$\Gamma$ is a subset of $\Lambda\times\Lambda$ and each pair
$(i,j)\in\Gamma$
is such that $i < j$. $ds_\Gamma\equiv \prod_{(i,j)\in\Gamma} \,ds_{ij}$,
$\partial^\Gamma = \prod_{(i,j)\in\Gamma} \frac{\partial}{\partial
s_{ij}}$,
$s_{\Gamma^c} = \{s_{ij}\}_{(i,j)\in\Gamma^c}$ and $|\Gamma|$ is the
cardinality of $\Gamma$. (19) then follows by writing
$1=\prod_{i<j}(I_{ij}+E_{ij})$ with
$I_{ij}=\int_{[01]}ds_{ij}{\partial\over\partial s_{ij}}$ and $E_{ij}$ is
the
evaluation map at $s_{ij}=0$ (for more details, see \cite{GJS} and
Proposition 18.2.2 in \cite{GJ}).

Now observe that $\delta_{\Lambda\Gamma}$ factorizes as follows. For a set
$A\subset\Lambda\times\Lambda$, let $\overline A$ denote the union of its
projections on the two factors. We say $p_1, p_2 \in\Gamma$ are connected
to
each other if $\overline p_1\cap \overline p_2\not= \emptyset$. Let
$\Gamma_\alpha$ be
the maximal connected components of $\Gamma$ with respect to this relation.
Then we have \be \delta_{\Lambda\Gamma} = \otimes_\alpha
\;\delta_{\overline\Gamma_\alpha\Gamma_\alpha} \otimes
1_{\Lambda\backslash\overline\Gamma} \en where
$\delta_{\overline\Gamma_\alpha
\Gamma_\alpha}: {\cal H}^{\overline\Gamma_\alpha}_{\rho_1} \to {\cal
H}^{\overline\Gamma_\alpha}_{\rho_1-\delta}$ is given by the integral in
(19)
(with $\Lambda$ replaced by $\overline\Gamma_\alpha$ and $\Gamma$ by
$\Gamma_\alpha$).

\vs{2mm}

\noindent We now estimate the norm of
$\delta_{\overline\Gamma_\alpha\Gamma_\alpha}$. Using (16) (which holds for
arbitrary $\Lambda$) and a Cauchy estimate in (19) (a circle of radius
$\frac{r}{2} e^{\frac{\lambda}{2} |i-j|}$ around any $s_{ij}$ in the
integral
(19) is inside $D_\Lambda$), we get \be \||\delta_{\overline\Gamma_\alpha
\Gamma_\alpha} \|| \leq \left(\frac{r}{2}\right)^{-|\Gamma_\alpha|}
e^{C\varepsilon r(1+ {\lambda}^{-d}) |\overline\Gamma_\alpha|} e^{-\frac
\lambda2
\sum_{(i,j)\in\Gamma_\alpha}|i-j|} . \en We arrive at the final formula
(11)
by setting \be \delta_Y = \sum_\Gamma \delta_{Y\Gamma} \en with $Y =
\overline\Gamma$ and $\Gamma$ connected. Given any such $\Gamma$ which we
may view
as a connected graph on the points of $Y$, pick a connected tree graph
$\tau$
with $\overline\tau = Y$ and estimate the sum over $\Gamma$ in (22) by \be
\||
\delta_Y\|| \leq (\sum_{j\in{\Bbb Z}^d} e^{-\frac12 {\lambda}|j|})^{|Y|}
\tilde\eta^{|Y| - 1}\sum_\tau e^{-\frac \lambda2 |\tau|} \en where
$\tilde\eta = r^{-1}\exp[C\varepsilon r(1+ {\lambda}^{-d})]$ (see Appendix
2
for more details). Since we may take $r\to\infty$ as $\varepsilon\to 0$,
see
(16-18), the claim (12) follows. \hfill$
\makebox[0mm]{\raisebox{0.5mm}[0mm][0mm]{\hspace*{5.6mm}$\sqcap$}}$ $
\sqcup$

\vs{3mm}

\no{\bf Remark.} Propositions 1 and 2 imply that there is an
$\varepsilon_0>0$ and $\rho>0$ such that ${\cal P}_\Lambda$ maps ${\cal
H}^\Lambda_\rho$ into itself for all $\Lambda$ and all
$\varepsilon\leq\varepsilon_0$: the domain of analyticity shrinks by an
amount $\delta$ when we apply $\Phi^\ast_\Lambda$, but it is expanded when
we
apply $P_\Lambda$ (by Proposition 1,a). So, we may choose $\varepsilon$
small
enough so that $\gamma\rho-\delta > \rho$. We will fix this $\rho$ now once
and for all.

\section{ Space time expansion for the invariant measure}

The $T_\Lambda$-invariant measure $\mu_\Lambda$ will be constructed by
studying ${\cal P}^n_\Lambda$ as $n\to\infty$. This will yield spectral
information on ${\cal P}_\Lambda$ uniformly in $\Lambda$. From $(8)$, we
have
\be {\cal P}^n_\Lambda = (\Phi^\ast_\Lambda P_\Lambda)^n, \en into which we
insert (11) for $\Phi^\ast_\Lambda$ and, for $P_\Lambda$, we use (5): \be
P_\Lambda = \otimes_{ i\in\Lambda} \;(Q_i + R_i) = \sum_{I\subset\Lambda}
\otimes_{i\in I}\; R_i\otimes_{ j\in\Lambda\backslash I}\; Q_j \equiv
\sum_{I\subset\Lambda} R_I\otimes Q_{\Lambda\backslash I}, \en and we get
\be
{\cal P}^n_\Lambda = \sum_{\{I_t\}}\; \sum_{\{{\cal Y}_t\}}\;\prod_{ t =
1}^n
(\otimes_{Y\in{\cal Y}_t} \; \delta_{Y}\otimes 1_{\Lambda\setminus {\cal
Y}_t}) (R_{I_t} \otimes Q_{\Lambda\setminus I_t} ) \en where the product of
operators is ordered, with $t=1$ on the right.

\vs{2mm}

\no To understand the structure of the terms in (26), let us first consider
a
simple example. Consider the term in (26) with ${\cal Y}_t=\emptyset$ for
$t\neq m<n$ and ${\cal Y}_m=\{Y\}$, where $Y$ is some finite subset of
${\bf
Z}^d$. Let also $I_t=\{i\}$ for $1\leq t\leq m$ and $I_t=\{j\}$ for
$m+1\leq
t\leq n$, where $i,j\in Y$. The contribution to (26) from this term is
given
by \be \left( (R^{n-m}_j\otimes Q_{Y\setminus j})\delta_Y (R^m_i\otimes
Q_{Y\setminus i} )\right )\otimes Q_{\Lambda\setminus Y} \en where we used
$Q^2=Q$ repeatedly. Let us write $Q$ in (6) as \be Q=h\hat{\otimes}\ell \en
where we identify bounded operators $ H_\rho\to H_\rho$ with elements of
$H_\rho$ tensored with its dual $ H_\rho^\ast$  and use $\hat{}$ to
distinguish this tensoring from the one used for spaces indexed by ${\bf
Z}^d$. Using the shorthand $$ \ell_Y\equiv \otimes_{i\in Y}\; \ell,\;h_Y =
\otimes_{i\in Y}\;h $$ and inserting (28) into (27), the latter becomes $$
W\otimes (h_{\Lambda\setminus j}\hat{\otimes}\ell_{\Lambda\setminus i}) $$
where $W:{\cal H}^{\{i\}}_\rho\to{\cal H}^{\{j\}}_\rho $ is given by (let
$g\in{\cal H}_\rho^{\{i\}}$) $$ Wg=\ell_{Y\setminus j}\left (R^{n-m}_j
\delta_Y(h_{Y\setminus i} \otimes R^m_ig)\right ). $$

\vs{2mm}

\no To make these remarks systematic, it is useful to introduce a
``space-time'' lattice ${\bf Z}^{d+1}$, where the extra dimension
corresponds
to the ``time'' $t$ in the product in (26), and to establish a
correspondence
between the terms in (26) and geometrical objects defined on this lattice.

\vs{2mm}

\no We let $\cal S$ denote the set of all finite ``spacelike'' subsets of
${\bf Z}^{d+1}$, i.e. $Z\in{\cal S}$ is of the form $Z=Y\times \{t\}$ for
some $Y\subset{\bf Z}^{d}$ and $t\in{\bf Z}$. Also, let $\cal B$ denote the
set of ``timelike bonds'' of ${\bf Z}^{d+1}$, i.e. $b\in\cal B$ is of the
form $b=\{(i,t),(i,t+1)\} \equiv b_i(t)$ for some $i\in {\bf Z}^{d}$ and
integer $t$. The correspondence with terms in (26) is defined as follows:
to
each $Y \in {\cal Y}_t , t = 1, \cdots, n$, we associate $Z=Y \times \{t\}$
and to each $I_t$ we associate the set of bonds $\{ b_i (t-1), i \in
I_t\}$.

\vs{2mm}

\no A {\it polymer} $\gamma$ is then defined as a connected finite subset
of
${\cal S}\cup{\cal B}$, i.e. the elements of $\gamma$ are spacelike subsets
and timelike bonds of ${\bf Z}^{d+1}$. We define two bonds $b,b'\in\gamma$
to
be connected, if $b\cap b'\neq\emptyset$ or if there exists a $Z\in\gamma$
such that $b$ and $b'$ intersect $Z$. $\gamma$ is then defined to be {\it
connected} if the set of bonds $b\in\gamma$ is connected with respect to
this
relation, or if $\gamma$ consists of a single element belonging to ${\cal
S}$. Denote by $\overline\gamma$ the support of $\gamma$, i.e. the subset
of ${\bf
Z}^{d+1}$ that is the union of the elements of $\gamma$. We say that two
polymers are {\it disjoint}, if $\overline{\gamma}_1\cap\overline{\gamma}_2
=\emptyset$. Thus, in the example above we have
$\gamma=\{Y\times\{m\},b_i(0),\dots, b_i(m-1),b_j(m), \dots ,b_j(n-1)\}$.

\vs{2mm}

\no Let $\gamma$ be a polymer. Denote by $\pi_s$ and $\pi_t$ the
projections
on ${\bf Z}^{d+1}={\bf Z}^{d}\times {\bf Z}$ to the first and the second
factor. Then $\pi_t(\overline{\gamma})$ is connected, i.e. it is an
interval
denoted $[t_-,t_+]$. Let $\gamma_\pm=
\pi_s(\overline{\gamma}\cap\pi_t^{-1}(t_\pm))$ The {\it weight} of the
polymer
$\gamma$, $W(\gamma)$, is a bounded linear operator $W(\gamma): {\cal
H}^{\gamma_-}_\rho\to{\cal H}^{\gamma_+}_\rho$ or equivalently
$W(\gamma)\in
{\cal H}^{\gamma_+}_\rho\hat {\otimes}({\cal H}^{\gamma_-}_\rho)^\ast$. $W$
in the example above is a weight with $\gamma_+=j$ and $\gamma_-=i$.

\vs{2mm}

\no We may now return to (26). Let us in general denote by
$\Lambda_{t_1t_2}$
the set $\Lambda_{t_1t_2} = \Lambda\times \{t_1,t_1 +
1,\dots,t_2\}\subset{\Bbb Z}^{d+1}$ where $-\infty\leq t_1 < t_2\leq\infty$
and $\Lambda\times \{t\}\equiv \Lambda_t$. $\gamma$ is said to be in
$\Lambda_{t_1t_2}$ if $\overline{\gamma} \subset \Lambda_{t_1t_2} $, and
each
$Z\in\gamma$ has $t_1+1\leq\pi_t(Z)\leq t_2$. In (26) we will encounter a
family of polymers $\Gamma$ in $\Lambda_{0n}$ and we need to make a
distinction whether their support intersects the boundary of $\Lambda_{0n}$
i.e. $\Lambda_{0}$ or $\Lambda_{n}$. We denote the family of those
intersecting neither by $\Gamma_v$ ($v$ stands for "vacuum polymers"),
those
intersecting $\Lambda_{0}$ but not $\Lambda_{n}$ by $\Gamma_0$, the ones
intersecting $\Lambda_{n}$ but not $\Lambda_{0}$ by $\Gamma_n$ and the ones
intersecting both by $\Gamma_{0n}$ (the example in (27) belongs to
$\Gamma_{0n}$). Finally, let $\partial\gamma$ be the set of $i\in
\overline\gamma$
such that $i$ belongs to exactly one $b\in\gamma$ and to no $Z\in\gamma$.

The convergence of the polymer expansion is due to two reasons: thanks to
(12), each $\delta_Y$ brings a small factor, which decays with the size of
$Y$ or the distance between the points of $Y$. On the other hand, bonds are
associated to $R$ factors and long strings of such factors are suppressed
by
(7). Note however that $\mu$ need not be small, only less than one. Thus,
this expansion is similar to the one of a lattice of weakly coupled one
dimensional systems, but where, within each system, the couplings are not
necessarily small.

\vs{2mm}

\no We have now

\vs{2mm}

\noindent {\bf Proposition 3.} (26){\it can be written as \be &&{\cal
P}^n_\Lambda = \sum_{\Gamma}\prod_{\gamma\in\Gamma_v} \langle
W(\gamma)\rangle\otimes_{\gamma\in\Gamma_{0n}}W(\gamma) \nonumber\\ &&
\otimes\left((h_{\Lambda_+}
\otimes_{\gamma\in\Gamma_n}W(\gamma)h)\hat{\otimes} (\ell_{\Lambda_-}
\otimes_{\gamma\in\Gamma_0}\ell W(\gamma))\right) \en where the sum is over
sets $\Gamma$ (possibly empty) of mutually disjoint polymers $\gamma$ with
$\partial\gamma\subset \Lambda_0\cup\Lambda_n$. The following notation was
used: $\ell W(\gamma)= \ell_{\gamma^+} W(\gamma)\in({\cal
H}^{\gamma_-}_\rho)^\ast\;$, $W(\gamma)h= W(\gamma)h_{\gamma_-}\in{\cal
H}^{\gamma_+}_\rho\;$, $\langle W(\gamma)\rangle= \ell_{\gamma^+}
W(\gamma)h_{\gamma_-}\in {\bf R}$ and $\Lambda_+=\Lambda
\backslash\cup_{\gamma\in\Gamma_{n} \cup \Gamma_{0n}} \gamma_+$ and
$\Lambda_-=\Lambda \backslash\cup_{\gamma\in\Gamma_{0} \cup \Gamma_{0n}}
\gamma_-$.

\no The weights satisfy \be \parallel \mid W(\gamma) \parallel \mid \leq
\mu^{B} \prod_{ Z\in\gamma} (C\eta)^{|Z|} T(Z, {\lambda}) \en where $B$ is
the number of bonds in $\gamma$ and $ \parallel \mid \cdot\parallel \mid $
is
the norm in ${\cal H}^{\gamma_+}_\rho\hat {\otimes}({\cal
H}^{\gamma_-}_\rho)^\ast$. }

\vs{2mm}

\no{\bf Proof.} Consider a given term in (26). $\{I_t\}_{t=1}^n$ determines
a
set of bonds ${\cal B}_0 =\{b_i(t)\;|\; i\in I_{t+1}\;,\; t\in[0,n-1]\}$ and
$\{{\cal Y}_t\}_{t=1}^n$ a subset of $\cal S$, ${\cal S}_0=
\{Y\times\{t\}\;,\; Y\in{\cal Y}_t\}$. Decompose ${\cal B}_0\cup{\cal
S}_0=\cup_{\gamma\in\Gamma}\gamma$ where $\Gamma$ is a set of mutually
disjoint polymers. Since $QR=RQ=0$, we see moreover that
$\partial\gamma\subset\Lambda_0 \cup\Lambda_n$. Thus the sum in (26) can be
written as \be {\cal P}^n_\Lambda = \sum_\Gamma O_\Gamma (\Lambda) \en
where
$\Gamma$ runs through such sets and $O_\Gamma (\Lambda) $ is the product in
(26) corresponding to $\Gamma$. Note also that we may, since $Q^2 = Q$,
replace each $Q$ in $O_\Gamma (\Lambda) $ which is on a bond that is
disjoint
from $\overline\Gamma$ and $\Lambda_0\cup \Lambda_n$ by the identity
operator. For
the remaining $Q$'s use the identity \be QKQ=\ell(Kh)Q, \en where $K:{\cal
H}_\rho\to{\cal H}_\rho$, to factorize $O_\Gamma ( \Lambda)$: We apply (32)
(extended to tensor products) to each product $Q_i\delta_Y Q_i$ in (26)
where
$i\in Y$. The result is the summand in (29) where the weights $W(\gamma):
{\cal H}^{\gamma_{\_}}_\rho\to{\cal H}^{\gamma_+}_\rho$ are given by (here
$F\in{\cal H}_\rho^{\gamma_{\_}}$) \be W(\gamma)F =
\ell_{Y\backslash\gamma_+} (O_\gamma(Y)(F\otimes
h_{Y\backslash\gamma_{\_}})). \en $Y$ in (33) is $\pi_s(\overline\gamma)$
and
$O_\gamma(Y)$ is given by the product in (26) with $\Lambda$ replaced with
$Y$.

\vs{2mm}

\no To bound $W(\gamma)$, we use (12) and (7), the bounds $\|\ell\|\leq 1$,
$\|h\|\leq C$ (the norm of linear functionals is also denoted by $\| \cdot
\|$) and the fact that on bonds $b$ of $ Y\times\pi_t(\overline\gamma)$
that are
disjoint from $\overline\gamma$, we have identity operators. \hfill$
\makebox[0mm]{\raisebox{0.5mm}[0mm][0mm]{\hspace*{5.6mm}$\sqcap$}}$ $
\sqcup$

\vs{2mm}

\no Equation (29) is an example of polymer expansion in statistical
mechanics
and it is well known (see e.g. \cite{B,P,S}) that the bounds (30) will
enable
us to prove exponential falloff of correlations (i.e. mixing) and construct
the $\Lambda\to{\bf Z}^d$ limit (in the standard treatments the weights of
polymers are scalars and not operators as here, but, as we will see, the
combinatorical part of the proof is as in the standard case). We refer the
reader not familiar with the combinatorical methods needed in this analysis
to the references cited above and to Appendix 2 and just spell out the main
steps here.

\vs{2mm}

\no First we wish to cancel in (29) the contribution from the polymers
$\gamma\in \Gamma_v$. Let us call these the vacuum polymers. These are
``freely floating'' in $\Lambda_{0n-1}$ unlike the others that are attached
to the boundary $\Lambda_{0}\cup\Lambda_{n}$, but they tend to cancel each
other. To see this, note that, by (10) with $G=1$, $\ell_\Lambda {\cal
P}^k_\Lambda = \ell_\Lambda$ for all $k$, so, since $\ell(h) = 1$, and
$Rh=RQh=0$ (which implies $W(\gamma)h=0$ for $\gamma\in \Gamma_0 \cup
\Gamma_{0n}$), we get \be 1 = \ell_\Lambda({\cal P}^{n-1}_\Lambda
h_\Lambda)
= \sum_{\Gamma}\;\prod_{\gamma} \langle W(\gamma)\rangle \en where $\Gamma$
is a set of disjoint vacuum polymers in $ \Lambda_{0n-1}$ (note that the
vacuum polymers in (29) lie in $\Lambda_{0 n-1}$ too). The cancellation we
are after is accomplished by using (34) to write \be {\cal
P}^n_\Lambda=(\sum_{\Gamma}\;\prod_{\gamma} \langle W(\gamma)\rangle )^{-1}
{\cal P}^n_\Lambda \en and substituting (29). The standard combinatorics
(see
Appendix 2) now yields

\vs{2mm}

\noindent {\bf Proposition 4.} (35) {\it can be written as \be {\cal
P}^n_\Lambda = \sum_\Gamma \otimes_{\gamma\in\Gamma_{0n}}V(\gamma) \otimes
\left((h_{\Lambda_+} \otimes_{\gamma\in\Gamma_n}V(\gamma)h)\hat{\otimes}
(\ell_{\Lambda_-} \otimes_{\gamma\in\Gamma_0}\ell V(\gamma))\right) \en
where
the sum is over sets $\Gamma$ (possibly empty) of disjoint polymers
$\gamma$
in $\Lambda_{0n}$ with
$\overline\gamma\cap(\Lambda_0\cup\Lambda_n)\neq\emptyset$ and
$\partial\gamma\subset \Lambda_0\cup\Lambda_n$. The weights satisfy} \be
\parallel \mid V(\gamma) \parallel \mid \leq \mu^{B} \prod_{ Z\in\gamma}
(C\eta)^{|Z|} T(Z, {\lambda}/2). \en

\vs{2mm}

\noindent Notice now that, as $n\to\infty$, the $\gamma\in{\Gamma}_{0n}$ in
(36) give exponentially small contributions due to (37), and (36) will
factorize. Let us define a function $\tilde{\Omega}_\Lambda\in {\cal
H}^\Lambda_\rho$ \be \tilde{\Omega}_\Lambda = \sum_{\Gamma}
h_{\Lambda\backslash\Gamma_+}\otimes_{\gamma\in\Gamma}V(\gamma)h \en where
$\Gamma$ is a set of finite, mutually disjoint polymers $\gamma$ in
$\Lambda_{-\infty 0}$ with $\overline\gamma\cap\Lambda_0\neq\emptyset$ and
$\partial\gamma\subset\Lambda_0$. Similarily, define a linear functional
$\widetilde{\cal L}_\Lambda:{\cal H}^\Lambda_\rho\to \Bbb R$ by \be
\tilde{\cal L}_\Lambda = \sum_{\Gamma} \ell_{\Lambda\backslash\Gamma_{\_}}
\otimes_{\gamma\in\Gamma} \ell V(\gamma) \en where $\gamma$ are now in
$\Lambda_{0 \infty}$, with $\emptyset\neq\partial\gamma\subset\Lambda_0$.
Put
\be \Omega_\Lambda = \frac{\widetilde\Omega_\Lambda}{\ell_\Lambda
(\widetilde\Omega_\Lambda)} ,\quad {\cal L}_\Lambda = \ell_\Lambda
(\widetilde\Omega_\Lambda) \widetilde{\cal L} _\Lambda. \en Then we have

\vs{2mm}

\noindent {\bf Proposition 5.} {\it (Spectral decomposition of
$P_\Lambda$).
There exist $\varepsilon_0 > 0$, $\mu_1 < 1$, $c < \infty$, independent of
$\Lambda$, such that, for $\varepsilon < \varepsilon_0$, $\Omega_\Lambda$
and
${\cal L}_\Lambda$ have the following properties \be &&
\|\Omega_\Lambda\|\leq e^{c|\Lambda|}, \;\;\;{\cal L}_\Lambda =
\ell_\Lambda
\\ && {\cal P}_\Lambda\Omega_\Lambda = \Omega_\Lambda \\ &&
\parallel\mid{\cal P}^n_\Lambda - \Omega_\Lambda\hat\otimes {\cal
L}_\Lambda
\parallel\mid \ \leq \mu^n_1\; e^{c|\Lambda|} \en for all
$\Lambda\subset{\Bbb Z}^d$ and $n\in\Bbb N$. }

\vs{2mm}

\noindent {\bf Remark.} Thus the spectrum of ${\cal P}_\Lambda$ consists of
the eigenvalue $1$ with multiplicity $1$ and the rest is in the disc
$\{|z|\leq \mu_1\}$, uniformly in $\Lambda$.

\vs{2mm}

\noindent {\bf Proof.} We have \be \ell_\Lambda (\widetilde\Omega_\Lambda)=
\sum_{\Gamma}\, \prod_\gamma\, \langle V(\gamma)\rangle \en and (using $\|
h
\| \leq C$) $$ \parallel\tilde{\Omega}_\Lambda\parallel\leq
C^{|\Lambda|}\sum_{\Gamma}\, \prod_\gamma\, \parallel V(\gamma)\parallel.
$$
Since $\parallel\ell\parallel\leq 1$, $|\langle V(\gamma)\rangle |$
satisfies
(37). Now standard estimates (see Appendix 2) give \be e^{-C|\Lambda|}\leq
\ell_\Lambda (\widetilde\Omega_\Lambda) \leq e^{C|\Lambda|},\;\;
\parallel\tilde{\Omega}_\Lambda\parallel\leq e^{C|\Lambda|} \en which
yields
the estimate in (41).

Consider next ${\cal P}^n_\Lambda - \widetilde\Omega_\Lambda \hat\otimes\;
\widetilde{\cal L}_\Lambda\ $. Using (36),(38) and (39), this is given
again
by (36), but with different constraints for the set $\Gamma$: either there
exists a $\gamma$ such that $\overline\gamma\cap \Lambda_0 \neq \emptyset,
\overline\gamma \cap \Lambda_n \neq\emptyset$ or these exists a pair
$\{\gamma,\gamma'\}$ such that $\overline\gamma\cap\overline\gamma\,' \neq
\emptyset$
and $\overline \gamma \cup \overline \gamma\,'$ intersects both $\Lambda_0$
and
$\Lambda_n$. The first set of terms come from ${\cal P}^n_\Lambda$ and are
not canceled by the corresponding terms in $\Omega_\Lambda \hat \otimes
{\cal
L}_\Lambda$ and the second set are the uncanceled terms coming from
(38,39).
In both cases there is an overall $\mu^n$ factor and using (37) and again
standard estimates for the combinatorics (see Appendix 2) we get \be
\parallel\mid{\cal P}^n_\Lambda - \Omega_\Lambda\hat\otimes {\cal
L}_\Lambda
\parallel\mid \leq \mu^n (1+C\eta)^n\, e^{c|\Lambda|}. \en Since $\eta\to
0$
as $\varepsilon\to 0$, (43) follows with $\mu_1=\mu (1+C\eta)$. To show
${\cal L}_\Lambda = \ell_\Lambda$ recall that, from (10), $\ell_\Lambda
{\cal
P}^n_\Lambda = \ell_\Lambda$ for all $n$. This together with (43) gives $$
\ell_\Lambda = \ell_\Lambda (\Omega_\Lambda\hat\otimes {\cal L}_\Lambda ) =
\ell_\Lambda(\Omega_\Lambda) {\cal L}_\Lambda ={ \cal L}_\Lambda. $$ To
prove
(42), use (43) to get $$ {\cal P}^n_\Lambda \Omega_\Lambda \longrightarrow
\ell_\Lambda (\Omega_\Lambda) \Omega_\Lambda = \Omega_\Lambda $$ as
$n\to\infty$, and then use the continuity of ${\cal P}_\Lambda$. \hfill$
\makebox[0mm]{\raisebox{0.5mm}[0mm][0mm]{\hspace*{5.6mm}$\sqcap$}}$ $
\sqcup$

\vs{2mm}

\noindent We will pass shortly to the $\Lambda = {\Bbb Z}^d$ limit but
before
that we need a more refined mixing condition than (43):

\vs{3mm}

\noindent {\bf Proposition 6.} {\it There exists $\mu_1 < 1, c < \infty$
such
that,  if $d \mu_\Lambda =  \Omega_\Lambda d m_\Lambda$ denotes the
$T_\Lambda$ invariant measure  constructed  in Proposition 5, then, for any
$X,Y \subset\Lambda$ and   $F\in{\cal H}^{X}_\rho$,$G\in{\cal H}^{Y}_\rho$,
\be
\big\vert\int(F\circ T^n_\Lambda) G\, d\mu_\Lambda - \int F\,
d\mu_\Lambda \int G\,d\mu_\Lambda\big\vert\leq  \mu^n_1\, e^{c \min
(|X|,|Y|)} e^{-\frac{\lambda}{4}d(X,Y)} \|F\|\;\|G\|
\en}

\vs{2mm}

\noindent {\bf Proof.} We want to compare \be \int (F\circ T^n_\Lambda)
G\,d\mu_\Lambda = \ell_\Lambda \left(F {\cal
P}^n_\Lambda(G\Omega_\Lambda)\right) \en with $$ \ell_\Lambda (F
\Omega_\Lambda) \ell_\Lambda (G\Omega_\Lambda). $$

\noindent To do this, we expand them. Let us denote the terms in the sum
(38)
by $\widetilde{\Omega}_\Lambda(\Gamma)$ and the ones in (39) by
$\widetilde{{\cal L}}_\Lambda(\Gamma)$. Then we get, see (40,41), \be
\ell_\Lambda(F \Omega_\Lambda) = \widetilde{\cal
L}_\Lambda(F\widetilde\Omega_\Lambda) = \sum_{\Gamma\Gamma'}
\widetilde{{\cal
L}}_\Lambda(\Gamma)F \widetilde{\Omega}_\Lambda(\Gamma') \en where $\Gamma$
is a set in $\Lambda_{0\infty}$, $\Gamma'$ in $\Lambda_{-\infty 0}$ and
their
members satisfy $\overline\gamma\,'\cap\Lambda_0\neq\emptyset$ and
$\emptyset\neq\partial\gamma\subset\Lambda_0$.

\noindent We want next to group together all $\gamma, \gamma'$ that
``connect'' to $X$. For this, consider the projections
$\pi_s(\overline\gamma)\equiv Y_\gamma $ of $\overline{\gamma}$ to
$\Lambda$ and
define $Y_{\gamma'}$ similarily. Define any two sets in
$\{X,Y_\gamma,Y_{\gamma'}\}$ to be connected if they intersect. Let $Y$ be
the connected component including $X$ under this relation and $\Gamma_Y$,
$\Gamma'_Y$ the sets of $\gamma$ and $\gamma'$ contributing to $Y$. We may
then rewrite (49) as \be \ell_\Lambda(F \Omega_\Lambda)=\sum_{Y: X\subset
Y}\ell_ {\Lambda \setminus Y}(\Omega_{\Lambda \setminus
Y})\sum_{\Gamma_Y,\Gamma'_Y} \widetilde{{\cal
L}}_Y(\Gamma_Y)F\widetilde{\Omega}_Y(\Gamma'_Y) \en  Since
$\ell_Z(\Omega_Z)=1$ for all $Z$, (50) equals \be \ell_\Lambda(F
\Omega_\Lambda)=\sum_\Gamma V_F(\Gamma) \en  where $\Gamma$ is a set of
polymers $\gamma$ in $\Lambda_{-\infty 0}$ or in  $\Lambda_{0\infty }$,
intersecting $\Lambda_0$ and with projections on $\Lambda$  connected to
$X$
in the above sense, and $V_F(\Gamma)$ is the corresponding term in  (51).
$\ell_\Lambda(G\Omega_\Lambda)$ has a similar expansion where the
$\Gamma$'s
are  connected to $Y$, so consider (48). We proceed as above, using (36):
\be
\ell_\Lambda \left(F {\cal P}^n_\Lambda(G\Omega_\Lambda)\right) = \sum_{
\overline\Gamma_1\cap \overline\Gamma_2 = \emptyset} V_F(\Gamma_1) V_G
(\Gamma_2) +
\sum_\Gamma V_{F,G} (\Gamma) \en where $\overline \Gamma = \bigcup_{\gamma
\in
\Gamma} \overline\gamma$, the first sum comes from the
$\Gamma_{0n}=\emptyset$
terms in (36) and, in the second one, we sum over sets $\Gamma$ with the
following properties:  $\Gamma=\Gamma_{-\infty 0}\cup
\Gamma_{0}\cup\Gamma_{n}\cup\Gamma_{0n} \cup\Gamma_{n\infty}$, with by now
an
obvious notation, and  $\Gamma_{0n}\neq\emptyset$; moreover, the
projections
$\pi_s(\gamma)$  are now connected to $X$ and $Y$. $V_{F,G}$ is
given by
\be
V_{F,G}(\Gamma)=\widetilde{{\cal L}}_Y(\Gamma_{n\infty}) F
(\otimes_{\gamma\in\Gamma_{0n}}V(\gamma)\otimes (
\widetilde{\Omega}_Y(\Gamma_{n})\hat{\otimes} \widetilde{{\cal
L}}_Y(\Gamma_{0})))G \widetilde{\Omega}_Y(\Gamma_{-\infty 0}).
\en
The bound
(47) follows now  from (51) and (53): the left hand side of (47) has an
expansion like (52) but  with $\overline\Gamma_1\cap \overline\Gamma_2\neq
\emptyset$ in the first sum and  then each term in both sums contributes
$\mu^n$  $e^{-\frac{\lambda}{4}d(X,Y)}$ (as in the proof of (43)). The
combinatorics is controlled by $e^{c \min (|X|,|Y|)}$ because  each
$\Gamma$
must be connected to $X$ and $Y$ (see Appendix 2).
 \hfill$
\makebox[0mm]{\raisebox{0.5mm}[0mm][0mm]{\hspace*{5.6mm}$\sqcap$}}$
$ \sqcup$

\section{ The $\Lambda \longrightarrow {\Bbb Z}^d$ limit: Proof of the
Theorem}

We have now all the ingredients for the proof of the Theorem. First we
describe the invariant measure $\mu$. \medskip \noindent To do this,
rewrite
$\widetilde\Omega_\Lambda$ in (38) as $$ \widetilde\Omega_\Lambda =
\sum_{\Gamma}\;\otimes_\gamma\; \widetilde V (\gamma) h_\Lambda $$ with
$\widetilde V(\gamma) = \frac{V(\gamma)h_{\gamma_-}}{h_{\gamma_+}}$. Since
$h
> 0$ on $S^1$ (Proposition 1), $h\not=0$ on $S_\rho$ for $\rho$ small
enough
and thus $\|\widetilde V\|$ satisfies (37) too. Thus we may exponentiate
(see Appendix 2) \be \widetilde\Omega_\Lambda = \exp(\sum_\gamma U
(\gamma))
h_\Lambda,  \en  $U$ satisfies \be \|U(\gamma)\|\leq \mu^{|B|}
\prod(C\eta)^{|Y|} T(Y, \frac{\lambda}{3}) \en  Define, for
$Y\subset\Lambda$, $H_Y\in {\cal H}^Y_\rho$ by $$  H_Y = \sum_{ \gamma_{+}
=
Y} U (\gamma) $$  where $\gamma\subset{\bf Z}^d_{-\infty0}$ and let
$H_{Y\Lambda}$ be given by  the same sum with
$\gamma\subset{\Lambda}_{-\infty0}$. Then  \be
\|H_Y\|,\;\|H_{Y\Lambda}\|\leq (C\eta)^{|Y|} T(Y,\frac{\lambda}{4})  \en
and
\be \|H_Y-H_{Y\Lambda}\|\leq (C\eta)^{|Y|} e^{-{\lambda\over 8}
d(Y,\Lambda^c)} T(Y,\frac{\lambda}{4}) \en where $d$ denotes the distance.
We
have $$ d\mu_\Lambda = \frac{\exp(\sum_{Y\subset\Lambda} H_{Y\Lambda})
h_\Lambda d m_\Lambda}{\int \exp ({\sum}_{Y\subset\Lambda} H_{Y\Lambda})
h_\Lambda\; dm_\Lambda} $$

\vs{2mm}

\no It is easy to see that there is a unique Gibbs measure corresponding to
the Hamiltonian $H$, for $\eta$ small. We take $\mu$ to be this measure,
i.e., the unique Borel probability measure on $M$ with conditional
probability densities given by \be \mu(dm_\Lambda\vert m_{\Lambda^c}) =
\frac{\exp[-\sum_{ Y\cap\Lambda\not=\emptyset} H_Y (m)]}{\int\ exp[-\sum_{
Y\cap\Lambda\not=\emptyset} H_Y (m)]} h_\Lambda dm_\Lambda \en for any
$\Lambda\subset{\Bbb Z}^d$ finite \cite{R,S}.

\vs{2mm}

\no The $\sigma_i$-invariance follows since the unique Gibbs measure
satisfying (58) is translation invariant. For the $T$-invariance of $\mu$
it
suffices to show $$ \int F\circ T\;d\mu = \int F\;d\mu $$ for all
$F\in{\cal
H}^{X}_\rho$, all $X$ finite. Let $T^{(\Lambda)} = T_\Lambda\;\otimes\;
f_{\Lambda^c}$ and $d\mu^{(\Lambda)} = d\mu_\Lambda\;\otimes\;h_{\Lambda^c}
dm_{\Lambda^c}$; then, $\mu^{(\Lambda)}$ is $T^{(\Lambda)}$ invariant.
Moreover $F\circ T$ is continuous and $F\circ T^{(\Lambda)}\rightarrow
F\circ
T$ in the sup norm. (56) and (57) imply that
$\mu^{(\Lambda)}\rightarrow\mu$
weakly (here $\Lambda\rightarrow{\Bbb Z}^d$ is taken in the sense of the
net
of finite subsets of ${\Bbb Z}^d$, see \cite{R,S}). Hence, $$ \int F\circ
T\;d\mu = \lim \int F\circ T^{(\Lambda)} d\mu^{(\Lambda)} = \int F\;d\mu.
$$
Mixing follows from (47) which carries over to the limit:

\vs{3mm}

\noindent {\bf Proposition 7.} {\it Let $\tau$ denote the ${\bf Z}^{d+1}$
action generated by $T$ and the $\sigma_i$. There exist $\alpha > 0$,
$c<\infty$, such that for all finite $X \subset {\bf Z}^d$, all $F$, $G\in
{\cal H}^{X}_\rho$, and all}  ${\bf n} \in {\bf Z}^{d+1}$
\be
\vert\int(F\circ \tau^{\bf n}) G\, d\mu - \int F\, d\mu \int G\,
d\mu\vert\leq e^{-\alpha |{\bf n}|} e^{c|X|} \|F\|\;\|G\| .
\nonumber
\en
\vs{3mm} \noindent Finally, to prove part 3 of the Theorem, it is enough,
by
a density argument, to show, for all finite $X \subset {\bf Z}^d$, and all
$F\in {\cal H}^{X}_\rho$ that
$$ \lim_{n\rightarrow\infty} \int F \circ T^n
dm= \int Fd\mu $$ But, $$ \int F \circ T^n dm = \lim_{\Lambda \to {\Bbb
Z}^d}
\int F \circ T^n_\Lambda dm =\lim_{\Lambda\to {\Bbb Z}^d} \ell_\Lambda
(F{\cal P}_\Lambda^n 1) $$ and $$ \int Fd\mu=
\lim_{\Lambda\rightarrow\infty}
\ell_\Lambda (F \Omega_\Lambda) $$ So, it is enough to show $$ \big\vert
\ell_\Lambda (F{\cal P}^n 1)- \ell_\Lambda (F\Omega_\Lambda) \big\vert\leq
\mu^n_1\, e^{c|X|} \|F\|\; $$ with $\mu_1$, $c$, independent of $\Lambda$.
But this  follows from (47) with $G=h^{-1}_\Lambda$. \hfill$
\makebox[0mm]{\raisebox{0.5mm}[0mm][0mm]{\hspace*{5.6mm}$\sqcap$}}$ $
\sqcup$

\vs{15mm}

\no{\Large\bf Acknowledgments}

\vs{15mm}

We would like to thank L.A. Bunimovich, E. J\"arvenp\"a\"a, J. Losson and
Y.G. Sinai $\;$ for interesting discussions. This work was supported by NSF
grant DMS-9205296 and by EC grants SC1-CT91-0695 and CHRX-CT93-0411.

\vs{15mm}

\no{\Large\bf Appendix 1: Proof of Proposition 1}

\vs{15mm}

\setcounter{equation}{0}

\no We work in terms of the lift of $f$ to the covering space $\Bbb R$ of
$S^1$, and  denote it by $f$ again. Thus, our assumptions are: There exists
a
$\rho_0 > 0$,  $\gamma > 1$ such that

\vs{2mm}

\no (a) $f$ is holomorphic and bounded on  $S_{\rho_0} = \{z\in{\Bbb
C}\big\vert |Im z| <\rho_0\}$ with  $f({\Bbb R}) = \Bbb R$,

\vs{2mm}

\no (b) $f(z+1) = f(z) + k\quad z\in S_{\rho_0}, \quad k\in{\Bbb N},\; k
\geq
2$,

\vs{2mm}

\no (c) $f^\prime (x) \geq \gamma\quad x\in\Bbb R$.

\vs{2mm} \setcounter{equation}{58} \noindent

Hence, by changing $\rho_0$, $\gamma$ if necessary, we may assume $
f(S_\rho)
\supset S_{\gamma\rho}\quad \rho\leq \rho_0 $ and $\psi \equiv f^{-1}$ is
holomorphic and bounded on $S_{\gamma \rho_0}$ and
$|f^\prime(z)|\geq\gamma$,  $|\psi'(z)|\leq 1/\gamma$ on $S_{\rho_0}$ and
$S_{\gamma\rho_0}$ respectively.  Let $H_\rho$ be the space of periodic
bounded holomorphic functions on $S_\rho$,  of period one (which can be
identified with $H_\rho$ of Sect.2). Since $P$ is  given by  \be (Pg)(z) =
\sum^{k-1}_{j=0}\; \frac{g\left(\psi(z+j)\right)}{f'\left(\psi(z+j)\right)}

\en  we get from the above remarks that $P:H_\rho\rightarrow
H_{\gamma\rho}$
for $\rho\leq \rho_0$, which is the claim a) of Proposition 1. \vs{2mm}

\noindent The proof of the spectral decomposition b) goes via finite rank
approximations. Let $\psi_q=f^{-q}$ and let $P_{qn}:H_{\rho}\to H_{\rho}$
be
the finite rank  operator $$
(P_{qn}g)(z)=\sum_{j=0}^{k^q-1}[(f^q)'(\psi_q(z+j))]^{-1}
\sum_{m=0}^{n-1}{1\over m!}g^{(m)}(\psi_q(j))(\psi_q(z+j)- \psi_q(j))^m .
$$
By Taylor's theorem and Cauchy's estimates, we have, for $z \in S_\rho$,
$Re
z \in [0,1]$, \be |P^qg(z)-P_{qn}g(z)|&\leq&{\gamma^{-q}\over n!}\sum_j
|\psi_q(z+j)- \psi_q(j)|^n\|g^{(n)}\|_{\rho/2}\nonumber\\ &\leq&
\gamma^{-q(n+1)}k^qC^n\rho^{-n}\|g\|_\rho \nonumber \en where we assume
$\psi_q (S_\rho) \subset S_{\rho/2}$, which holds for $q$ large enough and
we
used $\|\psi'_q\|_\rho\leq\gamma^{-q}$ in the second inequality. Since $g$
is
periodic, we may restrict ourselves to $Re z \in [0,1]$.

\vs{2mm}

Take now first $q$ large enough such that $C\rho^{-1}\gamma^{-q}
\leq{1\over
2}$ and then $n$ large enough such that $k^q2^{-n}< {1\over 2}$. Then $$
\parallel \mid P^q-P_{qn}\parallel \mid <{1\over 2} . $$ From this we
conclude, since $P_{qn}$ is of finite rank, that the spectrum of $P^q$
outside of the disc of radius $1\over 2$ consists of a finite number of
eigenvalues with finite multiplicities. Therefore the same holds for $P$
outside of a disc strictly inside the unit disc. On the other hand it is
well
known that in a space of $L^1$-functions of bounded variation, the spectrum
of $P$ consists of $1$ and a subset of $\{z\big\vert|z|\leq \mu < 1\}$ for
a
map like the one we are considering. Eigenvalue 1 comes with multiplicity 1
and the eigenvector $h$ is strictly positive \cite{Co}. Since eigenvectors
in
$H_\rho$ are also in $L^1$, we only need to prove that $h$ is in $H_\rho$.
If
this wasn't true, the spectral radius of $P$ in $H_\rho$ would be less than
1
and we would have $P^n1\to 0$ in $H_\rho$, hence in $L^1$, which is
impossible. \hfill$
\makebox[0mm]{\raisebox{0.5mm}[0mm][0mm]{\hspace*{5.6mm}$\sqcap$}}$ $
\sqcup$

\vs{15mm}

\no{\Large\bf Appendix 2: Combinatorics}

\vs{15mm}

\no

We collect here some details on the combinatorical estimates used in the
paper.\\ \vspace*{3mm} \par\noindent 1. {\bf Proof of (23):} \vspace*{2mm}
\par\noindent

Inserting (21) into (22), we get \be \parallel\mid \delta_Y \parallel\mid
\leq \tilde \eta^{|Y|-1} \sum_{\stackrel{\Gamma:}{\overline \Gamma = Y}}
e^{-
\frac{\lambda}{2} \sum_{(i,j)\in \Gamma}[i-j|} \en where we used
$|\overline
\Gamma| - 1 = |Y| - 1 \leq |\Gamma|$.\\ Now, associate to each $\Gamma$ in
(60) a tree graph $\tau = \tau (\Gamma)$ with $\overline \tau = Y$, and
write $$
\sum_\Gamma = \sum_\tau \sum_{\stackrel{\Gamma:}{\tau(\Gamma)=\tau}} $$ and
\be \sum_{(i,j) \in \Gamma} |i-j| \geq |\tau| + \sum_{(i,j) \in \Gamma
\backslash \tau} |i-j| \en Finally, the sum over $\Gamma$ with
$\tau(\Gamma)
= \tau$ is bounded by the  sum over all choices of lines connecting points
of
$Y$, which yields:  \be \sum_{\stackrel{\Gamma:}{\tau(\Gamma)=\tau}}
e^{-\frac{\lambda}{2}  \sum_{(i,j) \in \Gamma \backslash \tau}|i-j|}
 \leq \prod_{i \in Y} (1+ \sum_{\stackrel{j \neq i}{j \in Y}}  e^{-
\frac{\lambda}{2} |i-j|})
 \leq (\sum_{j \in \Bbb Z^d} e^{- \frac{\lambda}{2} |j|})^{|Y|}  \en and
this
proves (23).\hfill$
\makebox[0mm]{\raisebox{0.5mm}[0mm][0mm]{\hspace*{5.6mm}$\sqcap$}}$ $
\sqcup$
\\
\vspace*{3mm}
\par\noindent
2. {\bf Proof of Proposition 4:}

\vspace*{2mm} \par\noindent
Let us first consider the denominator in (35).
The basic result of the polymer expansion formalism \cite{B,P,S} is that
(34)
can be written as
\be
\sum_\Gamma \prod_\gamma \langle W(\gamma) \rangle =
\exp \sum_\gamma \tilde U (\gamma)
\en
whith
\be \tilde U (\gamma) =
\sum^\infty_{m=1} \frac{1}{m!} \sum_{\stackrel{(\gamma_1, \cdots,
\gamma_m)}{\cup^m_1 \overline{\gamma_i} = \overline \gamma}} \prod^m_{i=1}
\langle W
(\gamma_i) \rangle \sum_G \prod_{(i,j) \in G} \chi_{ij}
\en
 where we sum
over sequences of polymers in $\Lambda_{0n-1}$  (not necessarily disjoint
or
even distinct) and the sum $\sum_G$ is  over all connected graphs with
vertices $\{1,\cdots,m\}$ and $\chi_{ij}
 = 0$ if $\overline{\gamma_i} \cap \overline{\gamma_j} = \emptyset,
\chi_{ij} = -1$  if
$\overline{\gamma_i} \cap \overline{\gamma_j} \neq \emptyset$, so that this
sum
vanishes unless $\gamma$ is connected. $\tilde U (\gamma)$ satisfies the
bound:
\be
\vert \tilde U (\gamma) \vert \leq \mu^{B} \prod_{ Z\in\gamma}
(C\eta)^{|Z|} T(Z, {2\lambda}/3).
\en
Formulas (63, 64) follow from the polymer formalism, provided we have the
bound:
 \be
\sum_{\stackrel{\gamma:}{\overline{\gamma} \ni x}} | \langle W(\gamma)
\rangle |
\leq 	C \eta
\en
for any $x \in {\Bbb Z}^{d+1}$. To prove (66), we use (30) which holds also
for
$|\langle W(\gamma)\rangle|$, since $\| \ell \| \leq 1, \| h \| \leq C$.
Next,
note that
\be
\sum_{Z \ni x} (C \eta)^{|Z|} T (Z,\lambda) \leq \sum_{n
\geq 0}  \; \frac{(C \eta)^{n+1}}{n!} \sum_{(z_1,\cdots,z_n)} \sum_\tau
e^{-\frac{\lambda}{2}|\tau|}
\en
where $(z_1,\cdots,z_n)$ is a sequence of
mutually distinct $z_i  \in {\Bbb Z}^d$, and the last sum is over tree
graphs
on $\{x,z_1, \cdots,z_n\}$. Now, for each fixed tree, the sum over
$z_1,\cdots,z_n$ is bounded by $(C\lambda^{-d})^n$: we start by summing
over
vertices with incidence number one, remove those vertices, get a new tree
and
iterate (i.e., we ``roll back" the tree). The number of tree graphs is
bounded by $c^n n!$ so that (67) is bounded by $C \eta$.
\vs{2mm}

\noindent
{}From this we obtain (66) by repeating the argument in (62), so as to
reduce the sum in (66) to a sum over trees whose vertices are now sets $Z_i
\in S$,
and the edges carry powers of $\mu$. Then the sum over the trees is done as
in (67), using the fact that the edges are now one-dimensional (only in the
``time" direction) and that $\sum^\infty_{n=1} \mu^n = \frac{\mu}{1-\mu} <
\infty$. Of course we need to choose $\eta$, and therefore $\varepsilon$,
so
that $\frac{\eta}{1-\mu}$ is small enough. This finishes the proof of
(66), hence of (63,64).
\vs{2mm}

\noindent
To prove (65), we use (64) and sum first over the graphs $G$ corresponding
to a given tree
as in the proof of (62). Then one has to control the fact that, since
$\gamma$ in (64) can be written in many ways as a union of $\gamma_i$'s,
each
$Z$ in $\gamma$ can also be decomposed in many ways (each time-like bond in
$\gamma$ can also occur several times in the bonds of the $\gamma_i$'s, but
this
brings extra powers of $\mu$, and we use $\sum^\infty_{n=1} \mu^n <
\infty$).
Let us write $\eta =
\hat \eta (\eta / \hat \eta)$, and  $T(Z,\lambda) \leq
T(Z,\frac{\lambda}{3})T(Z,\frac{2\lambda}{3})$. For any term in (64), we
have
$$
\prod^n_{i=1} \prod_{Z \in \gamma_i} T(Z,\frac{2\lambda}{3}) \leq \prod_{Z
\in \gamma}
T (Z,\frac{2\lambda}{3})
$$
and the factor $\frac{1}{\hat \eta}$ can be absorbed into the constant $C$
in (65).
Then, the sum over all ways of decomposing $\gamma$ in (64) is bounded by
\be
\prod_{Z \in \gamma}
(\sum^\infty_{n=1}
\sum_{\begin{array}{c}(Z_i)^n_{i=1}\\ \cup Z_i=Z \end{array}} \prod^n_{i=1}
\hat \eta^{|Z_i|} T(Z_i,\frac{\lambda}{3}))
\leq \prod_{Z \in \gamma} \prod_{Y \subset Z} (1+
\sum^\infty_{n=1} (\hat \eta^{|Y|} T(Y,\frac{\lambda}{3}))^n)
\en
This in turn is bounded by
\be
\prod_{Z \in \gamma} \exp \left( \sum_{Y \subset Z}
\sum^\infty_{n=1} (\hat \eta^{|Y|} T (Y,\frac{\lambda}{3}))^n \right) \leq
\prod_{Z \in \gamma} \exp (C {\hat \eta} |Z|)
\en
using the fact that the
sum (67), and therefore each term in that sum, is bounded by $(C \eta)$;
this
holds for any $\lambda >0$, for $\eta$ small. So we may replace $\lambda$
by
$\frac{\lambda}{3}$ and $\eta$ by $\hat \eta$ and obtain (69). This
establishes (65).

\vs{2mm}

\noindent
Now, turning to the numerator in (35), write
\be
{\cal P}^n_\Lambda = \sum_\Gamma \varphi
(\Gamma) \sum_{\Gamma_v} \prod_{\gamma \in \Gamma_v} \langle W(\gamma)
\rangle
\en
where the sum over $\Gamma$ has the same constraints as in
Proposition 4, $\varphi (\Gamma)$ denotes the product in (29) (with
$\Gamma_v
= \emptyset)$ and the sum over $\Gamma_v$ runs over sets of disjoint vaccum
polymers in $\Lambda_{0n-1}$ so that $\overline{\gamma} \cap
\overline\gamma\,' =
\emptyset, \forall \gamma \in \Gamma, \forall \gamma' \in \Gamma_v$.
Applying
the polymer formalism to this last sum, we get
\be
\sum_{\Gamma_v}
\prod_{\gamma \in \Gamma_v} \langle W (\gamma) \rangle = \exp (
\sum_{\stackrel{\gamma:}{\overline \gamma \cap \overline \Gamma =
\emptyset}} \tilde U
(\gamma)),
\en
and, using (63), (70) becomes
\be
{\cal P}^n_\Lambda =
\sum_\Gamma \varphi (\Gamma) \exp (- \sum_{\stackrel{\gamma:}{\overline
\gamma
\cap \overline \Gamma \neq \emptyset}} \tilde U (\gamma))
\en
with $\overline \Gamma =
\bigcup_{\gamma \in \Gamma} \overline \gamma$. Now, we expand the
exponential
in (72), and combine each of the terms with $\Gamma$. Concretely, write
(72) as
$$ \sum_\Gamma \var
(\Gamma) \sum^\infty_{n=0} \frac{1}{n!} \sum_{(\gamma_i)^n_1} \prod^n_{i=1}
(- \tilde U (\gamma_i))
$$
where $\overline{\gamma_i} \cap \overline \Gamma \neq
\emptyset , \forall i.$ This can be written as
\be
\sum_\Gamma \var (\Gamma) \sum_{(n_\gamma)}
\prod_\gamma \frac{(-\tilde U (\gamma))^{n_\gamma}}{n_\gamma !} \en where
the
product runs over all polymers (the number of polymers is finite for $n$
and
$\Lambda$ finite) and $n_\gamma \in \Bbb N$. Now, decompose $$ \Gamma \cup
\{
\gamma | n_\gamma \neq 0\} = \cup \gamma'_i $$ into mutually disjoint
polymers and define $V(\gamma')$, for a polymer $\gamma'$, to be given by
the
sum (73), with the constraint
\be
\Gamma \cup \{ \gamma | n_\gamma \neq 0 \}
= {\gamma'} \en
Since $\tilde U (\gamma)$ is a number, this does not change
the type of operators being considered. Since the constraints on the sum
over
$\Gamma$ in (73) are the same as in (36), we have established (36).
\vs{2mm}

\noindent
The bound (37) follows from (30) applied to the factors in $\var (\Gamma)$
and to ${\tilde U} (\gamma)$. The sum over all possible decompositions of
$\gamma'$ in (74) can be controlled by using (68,69), and by going from
${2\lambda}/3$ to ${\lambda}/2$. \hfill$
\makebox[0mm]{\raisebox{0.5mm}[0mm][0mm]{\hspace*{5.6mm}$\sqcap$}}$ $
\sqcup$

\vspace*{3mm}
\par\noindent 3. {\bf Proof of Propositions 5 and 6}
\vspace*{2mm} \par\noindent
The inequalities (45) on $\ell_\Lambda (\tilde
\Omega_\Lambda)$ follow from (44) and the bounds (37) on $\parallel\mid
V(\gamma) \parallel\mid$, which hold also for $|\langle V(\gamma)
\rangle|$:
the polymer formalism allows us to write (44) as $\exp (\sum_\gamma {\hat
V}(\gamma))$, which is similar to (63,64) but with $\overline \gamma \cap
(\Lambda_0 \cup \Lambda_n) \neq \emptyset$. Then, a bound like (65) implies
(45).
\vs{2mm}

\noindent
To prove (46), consider the terms where $\Gamma \ni \gamma$, with
$\overline \gamma \cap \Lambda_0 \neq \emptyset,
\overline \gamma \cap \Lambda_n \neq
\emptyset$. Using (36,37), we bound them by
$$
\exp(c|\Lambda|)\sum_\gamma \mu^{B} \prod_{ Z\in\gamma} (C\eta)^{|Z|} T(Z,
{\lambda}/2)
$$
where $ e^{c|\Lambda|}$ controls the sum over $\Gamma
\backslash \gamma$ (as in (45)). Next, we extract a factor $\mu^n$ from
$\mu^{B}$ and we control the sum over $\gamma$ by
$(1+ c \eta)^n$ as follows: To each $\gamma$, associate a tree by
choosing, for each time $t=1,\cdots,n$, a set $Z_t$ (possibly empty) and a
time-like line joining successive non empty $Z$'s. The sum over the rest of
$\gamma$ is handled as in (62). The choice of the lines fixes
the ``origins", $x_t$ of $Z_t$. So, we have $|Z_t|$ choices for each line
and
$|\Lambda|$ choices for the intersection of $\gamma$ with $\Lambda_0$.
These
latter factors can be absorbed in $e^{c|\Lambda|}$ or $C^{|Z|}$ and $(1+c
\eta)^n$ is then an upper bound on
$$ \prod^n_{t=1} (1+ \sum_{\stackrel{Z
\subset \Lambda_t}{x_t \in Z}} (C \eta)^{|Z|} T (Z, \frac{\lambda}{2}))
$$
for any choice of $\{x_t\}$ (as in the bound on (67)).
The other terms contributing to the LHS of (46) are bounded in a similar
way.
\vs{2mm}

\noindent
The bound on (53) leading to (47) is also similar. We get, however,
$e^{c \min(|X|,|Y|)}$ instead of $e^{c|\Lambda|}$ (which is crucial)
because here all
contributing polymers are ``connected" to $X$ and $Y$. The connection is
defined
through the projection of $\overline \gamma$ on $\Lambda$, but it is easy
to see
that we may again reduce the estimates to a sum over tree graphs and that
for two
polymers $\gamma,\gamma'$ such that $\pi_s (\overline{\gamma}) \cap \pi_s
(\overline{\gamma}') \neq \emptyset$ and
$\overline \gamma \cap \Lambda_0 \neq \emptyset,
\overline \gamma\,' \cap \Lambda_0 \neq \emptyset$, we have that
\be
 d(\overline \gamma \,' , \overline \gamma)
\leq \sum_{Z \in \gamma'} [Z|.
\en
So, for $\gamma$ fixed, the sum over $\gamma'$ with
$\pi_s (\overline{\gamma})
\cap \pi_s (\overline{\gamma}') \neq \emptyset$ can be bounded by
$$
\sum_x
e^{-d(x,\overline \gamma)}\sum_{\overline{\gamma\,}' \ni x}
\mu^{B} \prod_{ Z\in\gamma'} (C\eta)^{|Z|} T(Z, {\lambda}/2)
$$
where we used
(75) to absorb a factor $e^{d(x,\overline \gamma)}$ into $\prod_{Z\in
\gamma'}
C^{|Z|}$. \\
Finally, in (55,
56, 57), we perform resummations, using (68,69), which explains why we have
smaller fractions of $\lambda$.


\newpage


\begin{thebibliography}{90} \bibitem{Bl} M.L. Blank, Periodicity and
periodicity on average in coupled map lattices. Rigorous results, preprint,
Observ. Nice. \bibitem{Bl2} M.L. Blank, Singular effects in chaotic
dynamical
systems, Russ. Acad. Dokl. Math. {\bf 47}, 1-5 (1993). \bibitem{B} D.C.
Brydges, A short course on cluster expansions, in: Critical Phenomena,
Random
Systems, Gauge Theories. Les Houches Session XLIII, K. Osterwalder, R.
Stora
eds., Elsevier, p.129-183 (1984). \bibitem{BC} L.A. Bunimovich, E. Carlen,
On
the problem of stability in lattice dynamical systems, preprint.
\bibitem{BS}
L.A. Bunimovich, Y.G. Sinai , Space-time chaos in coupled map lattices,
Nonlinearity, {\bf 1}, 491-516 (1988). \bibitem{BS2} L.A. Bunimovich, Y.G.
Sinai, Statistical mechanics of coupled map lattices, in Ref \cite{Ka}.
\bibitem{Bu} L.A. Bunimovich, Coupled map lattices: One step forward and
two
steps back, preprint (1993), to appear in the Proceedings of the ``Gran
Finale" on Chaos, Order and Patterns, Como (1993). \bibitem{Co} P. Collet,
Some ergodic properties of maps of the interval, lectures at the CIMPA
Summer
School ``Dynamical Systems and Frustrated Systems", to appear. \bibitem{ch}
M. Cross, P. Hohenberg, Rev.Mod.Phys. {\bf 65}, 851-1111 (1993).
\bibitem{DS} N. Dunford, J.T.
Schwartz, Linear Operators, Vol. 2, Interscience Publishers, New York, p.
1018 (1963). \bibitem{ER} J-P. Eckmann, D. Ruelle, Ergodic theory of chaos
and strange attractors, Rev. Mod. Phys. {\bf 57}, 617-656 (1985).
\bibitem{GJS} J. Glimm, A. Jaffe, T. Spencer, The particle structure of the
weakly coupled $P(\phi)_2$ model and other applications of high temperature
expansions, in: Contructive Quantum Field Theory, G. Velo, A.S. Wightman,
eds., Lectures Notes in Physics {\bf 25}, Springer, New York (1973).
\bibitem{GJ} J. Glimm, A. Jaffe, Quantum Physics. A functional integral
point
of view, Springer, New York (1981). \bibitem{Ka} K. Kaneko (ed): Theory and
Applications of Coupled Map Lattices, J. Wiley (1993). \bibitem{Ka2} Chaos:
Focus Issue on Coupled Map Lattices (ed. K. Kaneko), Chaos {\bf 2} (1993).
\bibitem{K} G. Keller, M. K\"unzle, Transfer Operators for coupled map
lattices, Erg. Th. and Dyn. Syst. {\bf 12}, 297-318 (1992). \bibitem{LM} A.
Lasota, M.C. Mackey, Chaos, Fractals and Noise, Springer, New York (1994).
\bibitem{LY} A. Lasota, J. Yorke, On the existence of invariant measures
for
piecewise monotonic transformations, Trans. Amer. Math. Soc. {\bf 186},
481-488 (1973).
\bibitem{MH} J. Miller, D.A. Huse, Macroscopic equilibrium from microscopic
irreversibility in a chaotic coupled-map lattice, Phys. Rev. E, {\bf 48},
2528-2535 (1993). \bibitem{PS} Y.G. Pesin, Y.G. Sinai, Space-time chaos in
chains of weakly coupled hyperbolic maps, in: Advances in Soviet
Mathematics,
Vol. 3, ed. Y.G. Sinai , Harwood (1991). \bibitem{P} C.E. Pfister, Large
deviations and phase separation in the two-dimensional Ising model, Helv.
Phys. Acta, {\bf 64}, 953-1054 (1991). \bibitem{Po} Y. Pomeau, Periodic
behaviour of cellular automata, J. Stat. Phys. {\bf 70}, 1379-1382 (1993).
\bibitem{R} D. Ruelle, Thermodynamic Formalism, Addison-Wesley (1978).
\bibitem{S} B. Simon, The Statistical Mechanics of Lattice Gases, Vol. 1,
Princeton Univ. Press (1994). \bibitem{Si} Y.G. Sinai, Gibbs measures in
ergodic theory, Russian Math. Surveys {\bf 27}, 21-64 (1972). \bibitem{Te}
R.
Temam, Infinite-Dimensional Dynamical Systems in Mechanics and Physics,
Springer, New York, 1988. \bibitem{V} D.L. Volevich, Kinetics of coupled
map
lattices, Nonlinearity {\bf 4}, 37-45 (1991). \bibitem{V2} D.L. Volevich,
The
Sinai -Bowen-Ruelle measure for a multidimensional lattice of interacting
hyperbolic mappings, Russ. Acad. Dokl. Math. {\bf 47}, 117-121 (1993).

\end{thebibliography}
\end{document}